\documentclass{article}

\usepackage{arxiv_2}

\usepackage[utf8]{inputenc} 
\usepackage[T1]{fontenc}    
\usepackage{hyperref}       
\usepackage{url}            
\usepackage{booktabs}       
\usepackage{amsfonts}       
\usepackage{nicefrac}       
\usepackage{microtype}      
\usepackage{lipsum}		
\usepackage{graphicx}
\usepackage{natbib}
\usepackage{doi}

\usepackage{multirow}
\RequirePackage{natbib}
\hypersetup{colorlinks,allcolors=blue}
\renewcommand{\shorttitle}{Improved DR Estimator Using Partially Recovered Unmeasured Spatial Confounder}

\RequirePackage{amsthm,amsmath,amsfonts,amssymb}
\RequirePackage{float}

\newcommand\ci{\perp\!\!\!\perp}

\newcommand\bs{\boldsymbol{s}}
\newcommand\bY{\boldsymbol{Y}}
\newcommand\bA{\boldsymbol{A}}
\newcommand\bU{\boldsymbol{U}}
\newcommand\bZ{\boldsymbol{Z}}

 \def \bgamma{\boldsymbol{\gamma}}
 
 \def \blambda{\boldsymbol{\lambda}}
 \def \bSigma{\boldsymbol{\Sigma}}
 \def \bmu{\boldsymbol{\mu}}
 \def \E{\mathbb{E}}
 
\begin{document}

\title{An Improved Doubly Robust Estimator Using Partially Recovered Unmeasured Spatial Confounder}

\author{Sayli Pokal$^\ast$\\[4pt]
\textit{University of Nebraska-Lincoln, Nebraska, USA}
\\[2pt]
Yawen Guan\\[4pt]
\textit{University of Nebraska-Lincoln, Nebraska, USA}
\\[2pt]
Honglang Wang\\[4pt]
\textit{Indiana University–Purdue University Indianapolis, Indiana, USA}
\\[2pt]
Yuzhen Zhou\\[4pt]
\textit{University of Nebraska-Lincoln, Nebraska, USA}
\\[2pt]
{saylipokal18@gmail.com}
}

\markboth{}{}

\maketitle


\begin{abstract}
{Studies in environmental and epidemiological sciences are often spatially
varying and observational in nature with the aim of establishing cause and effect relationships. One of the major challenges with such studies is the presence of unmeasured spatial confounders. ‘spatial confounding’ is the phenomenon
in which the spatial residuals are correlated to the spatial covariates in the
model, when unaccounted for it can lead to biased causal effect estimates. This paper develops a novel method that adjusts for the
spatial confounding bias under a spatial-causal inference framework when
the treatment is binary. By combining tools from spatial statistics and causal
inference literature, we propose a method that reduces the bias due to spatial
confounding. We partially recover the unmeasured spatial confounder using
the spatial residuals and propose an improved doubly robust estimator based
on it. Through simulation studies, we demonstrate that the proposed doubly
robust estimator outperforms the existing methods and has the lowest bias
and close to nominal coverage in most scenarios. Finally, we implement our method to estimate the effect of installing SCR/SNCR $NO_x$ emission control technologies on ambient ozone concentrations.}


\end{abstract}


\section{Introduction}

Studies in environmental and epidemiological sciences are often observational, and the data in these studies may vary geographically. When the goal of such observational studies is to establish causal relationships between the treatment and the outcome, one of the major challenges is the presence of unmeasured confounders. For example, the impact of air pollution regulations on the ambient air quality and the corresponding effect on human health outcomes is constantly monitored \citep{zigler2016causal, zigler2018impact}. Unmeasured spatial confounding exists when variables such as unknown pollutants related to both the treatment and the outcome vary geographically and are not observed. Unmeasured spatial confounding when unaccounted for can result in biased causal effect estimates. When the data at hand is both observational and spatial, methods that combine the knowledge of causal inference methodology with spatial statistics are needed.

In practice, regression models with a spatial random effect term are often employed to `account' for unmeasured spatial confounders. Such models assume the covariates are independent of the spatial residuals. However, in the presence of an unmeasured spatial confounder, the spatial residuals are collinear to the treatment variable, resulting in biased causal effect estimates. Studies have shown that including a spatially correlated random effect term does not eliminate the bias caused by spatial confounding \citep{paciorek_2010, hodges_reich_2010}. \cite{hodges_reich_2010} considered the implications of adding a spatial random effect term under different settings and suggested using restricted spatial regression to avoid spatial confounding. \cite{paciorek_2010} showed that the bias due to spatial confounding is substantial in many scenarios and cannot be eliminated. Further, he reported that inclusion of a spatial residual term helps reduce the spatial confounding bias ``only when there is unconfounded variability in the exposure at a scale smaller than the scale of confounding''. These methods in the spatial statistics literature focus on adjusting the outcome regression models to account for unmeasured spatial confounding, but do not target estimation of causal effects as such.   

In the spatial causal inference literature, methods have recently been developed to adjust for the bias caused by missing spatial confounders. \cite{schnell_2019_mitigating} proposed an affine estimator to mitigate the spatial confounding bias. They provided a set of assumptions on the joint distribution of the missing confounder and the treatment variable such that the unmeasured spatial confounding bias term can be identified and corrected. However, it might be difficult to verify if the assumptions on the joint distribution hold in practice.
Under a regression discontinuity design, \cite{keele_enhancing_2015} matched observational units in the treatment and control areas to achieve covariate balance and minimize the distance between matched pairs. 
Similarly, \cite{dapsm} proposed a distance adjusted propensity score matching (DAPSm) method.
The method provides matches using a distance-adjusted propensity score, a weighted average of the propensity score differences, and the geographic distance between the treated-control pairs. 
Additional methods to adjust for spatial confounding are as follows, \cite{thaden2018structural} used structural equation modeling to adjust for unmeasured spatial confounders in geoadditive models. \cite{keller_2019_selecting} adjusted for spatial confounding bias using splines with Fourier and wavelet filtering. In the causal inference literature, instrumental variables are often used to adjust for unobserved confounding variables. \cite{giffin_instrumental_2021} explored the effectiveness of the instrumental variable approach in the presence of unmeasured spatial confounding. They found that the spatial instrumental variables are most effective if they vary at a finer spatial resolution than the spatially varying continuous treatment. Spatial causal inference is a growing research area; see \cite{reich_review_2020} for a complete review of methods that have recently been developed in this field.

In this paper, we propose a novel method to adjust for unmeasured spatial confounding in the continuous (geostatistical) spatial domain when the treatment is binary, and the outcome is continuous. In the spatial literature, treatment is often considered to be continuous \citep{paciorek_2010, schnell_2019_mitigating, giffin_instrumental_2021, keller_2019_selecting}, whereas, in causal inference, it is common to have a binary treatment. The existing literature on methods that account for unmeasured spatial confounding in the case of a binary treatment in geostatistical data is limited. To our knowledge, it is restricted to \cite{dapsm} and \cite{ keele_enhancing_2015} which are based on the idea of matching and using geographic distance as a proxy to adjust for unmeasured spatial confounders. However, matching observational units based on geographic proximity might result in many unmatched units being discarded. Discarding unmatched units changes the population on which inference is made and makes it difficult to interpret the estimated causal effects. 
Additionally, the distance-adjusted propensity score method computes causal effect estimates using only the propensity score model. If the propensity score model is misspecified, the causal estimates from this method may suffer from additional bias due to model misspecification.

The goal of this study is to obtain estimates of causal effects that reduce the bias due to unmeasured spatial confounding. We achieve this by improving the propensity score estimates so that balance can be achieved on both measured and unmeasured confounders. 
We propose an alternate way to incorporate spatial information in the propensity score model. 
Unlike the existing methods that use geographic proximity to adjust for unmeasured spatial confounders, our method uses spatial residuals to extract information about the missing spatial confounders. Thus, our proposed method does not require computing proximity by matching observational units. 
Furthermore, adjusting for spatial confounding using the spatial dependence structure of the missing confounder can provide improved causal estimates compared to simply using the geographic proximity of units. 
In addition to improving the propensity score estimation, our method uses a doubly robust estimator to obtain causal effect estimates, which are robust to model misspecification when either model is misspecified. We show that in most scenarios, our proposed method has the lowest bias and the lowest variance compared to the existing methods through simulation studies. 

In the case of binary treatment and continuous outcome in geostatistical data, our proposed method can be used to obtain the most common causal effect estimates of average treatment effect (ATE), average treatment effect on the treated (ATT), or average treatment effect on the control (ATC) \citep{imbens2015causal}. 
While in this paper we focus on estimating ATT, our proposed method can be easily generalized to estimate ATE or ATC. The rest of the paper is organized as follows; in Section \ref{Sec: spatial causal inference framework}, we set up the problem of spatial confounding under the spatial causal inference framework. In Section \ref{Sec: Adjusting for unmeasured spatial confounders}, we introduce our proposed method to adjust for unmeasured spatial confounders. In Section \ref{Sec: Simulation}, we conduct an extensive simulation study to assess the performance of our method and compare it with alternative methods. In Section \ref{Sec: Case study}, we analyze the power plant emissions data set and study the effect of installing selective catalytic reduction or selective non-catalytic reduction (SCR/SNCR) technologies on ambient ozone compared to alternatives. We end with discussions in Section \ref{Sec: Discussions}.   
\section{Spatial causal inference framework under a binary treatment}
\label{Sec: spatial causal inference framework}
\subsection{Spatial confounding}
Let $\{Y(\bs),\ \bs \in \mathcal{D} \subset \mathbb{R}^d\}$ be a continuous spatial process, in the domain of interest $\mathcal{D}$. At every location $\bs \in \mathcal{D}$, suppose that there exists a binary treatment $A(\bs)$ and $p$ observed confounders $\bZ(\bs) = (Z_1(\bs), ..., Z_p(\bs))^\top$. We further assume that the spatial process $U(\bs)$ represents all the missing spatial confounders; $U(\bs)$ is correlated to both $A(\bs)$ and $Y(\bs)$. 
Consider the following spatial regression model, 
\begin{align}
    Y(\bs) = A(\bs)\beta + \bZ(\bs)^\top \bgamma + U(\bs)  + \epsilon(\bs), \label{eq:outcome}
\end{align}
where $\epsilon(\bs)$ is an independent Gaussian process with mean zero and variance $\sigma_e^2$. Suppose that data were observed on a finite collection $\mathcal{S} = \{\bs_1, ..., \bs_n\}$. Let $\bY = \ \  (Y(\bs_1), ..., Y(\bs_n))^\top$, $\bA = (A(\bs_1), ..., A(\bs_n))^\top$, $\bU = (U(\bs_1), ..., U(\bs_n))^\top$ and $\bZ = (\bZ(\bs_1), ..., \bZ(\bs_n))^\top$. For notation convenience, we also use $Y_i$, $\bZ_i$, $A_i$, and $U_i$ to denote the observed $Y$, $\bZ$, $U$, $A$ at location $\bs_i$.


A common approach to account for unmeasured spatial confounding in spatial statistics is by modeling $\bU$ as a spatial random effect term following the multivariate Gaussian distribution. Specifically, $\bU \sim Gaussian(0, \bSigma_U)$, where $\bSigma_U$ is a spatial covariance matrix. The generalized least squares (GLS) estimate of $\beta$ under unmeasured spatial confounding follows,
\begin{align}
        &\E(\hat \beta_{GLS}\,|\, \bA) = \beta + \big(\bA^\top (\bSigma_U+\sigma_e^2 \mathbf{I})^{-1} \bA\big)^{-1} \bA^\top (\bSigma_U + \sigma_e^2\mathbf{I})^{-1} \E(\bU\,|\,\bA), \label{bias}
\end{align} 
where the second term denotes the bias. \cite{paciorek_2010} derived the explicit form of the bias in the GLS estimate of $\beta$ in case of continuous treatment, which showed that in the presence of spatial confounding, including a spatial random effect term does not eliminate the bias. However, the bias cannot be derived analytically for a binary treatment $\bA$.  The impact of adding a spatial random effect term is not well understood and does not always lead to a reduction in bias. Moreover, the approach does not focus on estimation of causal effect.


\subsection{Causal inference framework}

In a randomized controlled trial (RCT), since the treatment is randomized, the confounders are balanced, i.e., the distribution of the confounding variables is similar in the treatment and the control groups. 
In the case of observational data, treatment is not randomized; the confounders are not balanced across the treatment and control groups. Thus, there is a need to adjust for the confounders to obtain valid causal effect estimates.
In this study, we will adopt the potential outcomes framework, also known as the Neyman-Rubin causal model \citep{rubin1974estimating}, to identify the causal effect estimates in spatial statistics. 


\subsubsection{Potential outcomes framework and causal assumptions}

Under the potential outcomes framework, each observational unit is assumed to have two potential outcomes. Recall that $A(\bs)$ denotes the treatment received at the location $\bs$, $A(\bs) = 1$, if the unit at $\bs$ is treated and $A(\bs) = 0$ if untreated. Let $Y^{(1)}(\bs)$ and $Y^{(0)}(\bs)$ denote the two potential outcomes corresponding to whether the unit at $\bs$ is treated or untreated.

We make the stable unit treatment value assumption (SUTVA) which states that there is a single version of the treatment and that the treatment at one spatial location does not affect the outcome at other spatial locations (no interference). In spatial settings, such assumption is valid when the spatial locations are far enough or when the treatment has a local effect causing no interference between neighboring locations. Let us assume for now that $\bU$ is observed. The no unmeasured confounders assumption states that, the conditional probability of receiving every value of treatment at any location $\bs$, depends \textit{only} on the measured covariates $(\bZ(\bs),U(\bs))$, i.e.,  $( Y^{(1)}(\bs), Y^{(0)}(\bs) ) \ci A(\bs)\, |\, (\bZ(\bs), U(\bs))$. Further, we assume that the causal assumptions of positivity and consistency hold. 


\subsubsection{Adjusting for observed confounders}
If $\bU$ is observed and the causal assumptions are met, the treatment effect estimate from the outcome regression model in \eqref{eq:outcome} provides a valid causal estimate if it is correctly specified. 
Alternatively, if the outcome model is unknown,  methods based on propensity scores can be used to adjust for confounding and to estimate causal effects from observational data. 
For any location $\bs$, the propensity score is defined as, $e\big(\bZ(\bs), U(\bs)\big) := P\big(A(\bs)=1\,|\,\bZ(\bs), U(\bs)\big)$.
The true propensity score is unknown and for a binary treatment it is commonly estimated using logistic regression. 
Popular methods based on propensity score include, propensity score matching, inverse probability treatment weighting, and doubly robust estimators \citep{austin2011introduction}. 

In this paper, we are interested in estimating ATT. 
It can be estimated from the observed data using the inverse probability treatment weighting (IPTW) estimator as follows,
\begin{align}
ATT_{IPTW} = \frac{1}{n} \sum_{i=1}^{n} {A_i Y_i} - \frac{1}{n} \sum_{i=1}^{n} \frac{e(\bZ_i, U_i) (1-A_i)  Y_i}{ 1 - e(\bZ_i, \bU_i)} \label{eq:att_iptw}
\end{align}
If the propensity score model is correctly specified, the IPTW estimator provides an unbiased estimate of ATT \citep{lunceford2004stratification}.  
Another popular estimator is the doubly robust (DR) estimator, given by,
   \begin{align}
     ATT_{DR} &=   
     \frac{\sum_{i=1}^{n} \big( A_i - (1- A_i) e(\bZ_i, U_i) \big) \big(Y_i - \mu_i^{(0)} (\bZ, \bU) \big)  }
     {\sum_{i=1}^{n} A_i}, \label{eq:att_dr}
     \end{align}
where $\bmu^{(0)}(\bZ, U) := \E(\bY|\bA=0, \bZ, \bU)$ is the conditional mean for $\bY$ in the control population. DR estimator offers protection against misspecification when either outcome model or propensity score model is misspecified \citep{bang2005doubly}.

Assuming that $\bU$ is observed and all causal assumptions are satisfied, valid causal effect estimates can be obtained from the observed data using the standard causal inference methods described above. However, when $\bU$ is unobserved, it violates the causal assumption of no unmeasured confounders; the causal effect estimates obtained using the above methods will be biased. Since in our case the unmeasured confounder also has a spatial structure, it calls for methods that adjust for spatial confounding to reduce bias in treatment effect estimates. In the next section, we propose a novel method to adjust for the unmeasured spatial confounders.

\section{Adjusting for unmeasured spatial confounders using RecoverU} 
\label{Sec: Adjusting for unmeasured spatial confounders}

The aim of this paper is to reduce the bias in causal estimates in the presence of unmeasured spatial confounding in a geostatistical data setting with binary treatment. 
Suppose that the data is generated from the following model,
 \begin{align}
    &Y(\bs) = A(\bs)\beta + \bZ(\bs)^\top \bgamma + U(\bs)  + \epsilon(\bs), \label{eq:1} \\ 
    &A(\bs) \sim \ Bernoulli (e(\bZ(\bs),U(\bs))),  \\
    &e(\bZ(\bs),U(\bs)) = expit(\bZ(\bs)^\top\blambda + U(\bs) \alpha), \label{eq:3}
\end{align}
where the coefficients $\beta, \alpha \in \mathbb{R}$, $\bgamma, \blambda \in \mathbb{R}^p$, the function $expit(x):= 1/(1+e^{-x})$, and $\epsilon(\bs)$ is an independent Gaussian process with mean zero and constant variance $\sigma_e^2$. We are interested in estimating the treatment effect $\beta$, which is a scalar.

If $\bU$ is unobserved and is correlated to $\bA$, some of the variability in $\bU$ gets attributed to $\bA$. Approaches such as least squares estimation and maximum likelihood estimation favor attribution of variability in $\bU$ to the fixed effect term $\bA$ to minimize the residual sum of squares or maximize the likelihood; this leads to bias in the treatment effect estimates. Our proposed method recovers spatial residuals from the outcome model for the purpose of improving the propensity score estimates and then uses tools in causal inference to obtain causal estimates. The method can be broken down into three steps; recovering the unmeasured confounder, estimating propensity scores and obtaining the doubly robust causal estimates. 


\subsection{Recovering the unmeasured confounder}
The variability in $\bU$ can be decomposed to, \(\bU_A \text{ and } \bU_R, \text{ where } \bU_R \ci \bA \text{ and } \bU_A = \E(\bU\,|\, \bA) \). $\bU_A$ is the variability in $\bU$ attributed to $\bA$, and $\bU_R$ is orthogonal to $\bA$. With the classical spatial regression, which assumes the spatial random effect is independent of covariates, $\bU_A$ cannot be recovered.  But we are able to approximately recover $\bU_R$. Here are the detail steps.
\begin{enumerate}
    \item Assume that the random effect $U(\cdot)$ in the spatial regression model \eqref{eq:1} is a centered Gaussian random field with Mat\'ern covariance function, i.e., 
    \begin{align}
        C(h;\sigma^2, \theta, \nu) = \frac{\sigma^2}{\Gamma(\nu) 2^{\nu - 1} }
    \left( \frac{2 \sqrt{\nu} h }{\theta} \right)^{\nu}
    \kappa_{\nu} \left( \frac{2 \sqrt{\nu} h }{\theta} \right), 
    \end{align}
    where $h$ is the Euclidean distance between two locations, $\sigma^2$ is the variance parameter, $\nu$ is the spatial smoothness parameter, $\theta$ is the spatial range parameter, and $\kappa_{\nu} (\cdot)$ is the modified Bessel function of the second kind \citep{stein1999interpolation}. Using the iteratively re-weighted least squares (IRWLS) approach, we obtain the estimates of the covariance parameters $(\sigma^2, \nu, \theta)$ and the nugget variance $\sigma_e^2$, say, $(\hat\sigma^2, \hat\nu, \hat\theta, \hat \sigma_e^2)$; accordingly, we got the estimated covariance matrix of $\bU$, i.e., $\hat\bSigma_U$.
    \item Obtain the generalized least squares (GLS) estimates of $\beta$ and $\bgamma$, say, $\hat \beta$ and $\hat\bgamma$. Due to the spatial confounding, $\hat \beta$ is biased. Indeed, the bias term $\bA(\hat\beta-\beta)$ provides an estimate of $\bU_A$. Yet, we are not able to recover $\bU_A$ with $\beta$ unknown.
    \item Note that $\bA \hat \beta$ approximately recovers $\bA\beta + \bU_A$. We can recover $\bU_R$ via the spatial residuals $\mathbf \bY - \bA \hat\beta - \bZ\hat\bgamma$. Specifically, 
\begin{align}
  \hat \bU_R := \hat\bSigma_U [\hat\bSigma_U + \hat\sigma_e^2 \mathbf{I}]^{-1} ( \mathbf \bY - \bA \hat\beta - \bZ\hat\bgamma).\label{eq:4}
\end{align}
\end{enumerate}

\subsection{Propensity score estimates}
The recovered component of $\bU$, $\hat \bU_R$ can be considered an approximation of the true $\bU_R$ and treated as an observed covariate. 
The true propensity scores, $e(\bZ_i, U_i)$ in \eqref{eq:3} can be approximately estimated by fitting the following model, 
\begin{align}
   &A_i \sim \ \text{Bernoulli} (e(\bZ_i, \hat U_{R,i})),  \\
&e(\bZ_i,  \hat U_{R,i}) =  expit( \bZ_i^\top\blambda + \alpha \hat U_{R,i}), \ i = 1, 2, ..., n.
\end{align}
The propensity score estimates obtained by including $\hat \bU_R$ in the propensity score model will achieve balance on all observed confounders while also adjusting for the unmeasured $\bU$. Unlike the methods that adjust for spatial confounders using geographic proximity of the observations, our method provides a way to assess the covariate balance and overlap between treated and control groups using $\hat \bU_R$.

\subsection{Causal effect estimates}

Our method improves the propensity score estimates by incorporating information about the missing confounders in the propensity score model. Thus, any propensity score-based method such as matching, IPTW, or DR estimators can be used to obtain causal effect estimates of ATE, ATT, or ATC, which will be less biased than the estimates that do not adjust for the missing confounder.
We propose using DR estimators to obtain causal estimates as it offers protection against model misspecification. If the propensity score model is correctly specified, the causal estimates obtained using DR estimators have a smaller variance than IPTW estimators. If both outcome and propensity score models are correctly specified, these estimators will have the lowest mean squared errors (MSE) \citep{lunceford2004stratification}. In this paper, we focus on estimating the ATT, which is given by \citep{moodie2018doubly}, 
\begin{equation}
ATT_{DR} =   
\frac{\sum_{i=1}^{n} \big( A_i - (1- A_i) \hat e(\bZ_i, \hat\bU_{R,i}) \big) \big(Y_i - \hat \mu^{(0)} (\bZ_i, \hat\bU_{R,i}) \big)  }
{\sum_{i=1}^{n} A_i}, \label{eq:att_dr_uhat}
\end{equation}
where $\hat e(\bZ_i, \hat \bU_{R,i})$ is the estimated propensity score and $\hat \mu^{(0)}(\bZ_i, \hat \bU_{R,i} ) = \bZ_i^\top \hat \gamma + \hat \bU_{R,i}$. The variance estimate for the doubly robust estimator is obtained by bootstrapping.
\section{Simulation study} 
\label{Sec: Simulation}
We conducted a series of simulation studies to examine our proposed method's performance compared to existing alternatives under different scenarios. The objectives of the study were to (1) assess how the proposed method improves propensity score estimation; (2) examine the impact of the strength of spatial confounding on the causal estimates; (3) explore how changes in the spatial structure of the unmeasured confounder impact the estimation of the causal effects; and (4) assess the robustness of the different methods under model misspecification.       

\subsection{Data} \label{simulated data}

For a fixed set of 500 locations in a 2D space, we generated the unmeasured spatial confounder $\bU$ to be a Gaussian random field with mean zero and Mat\'ern covariance function $C(h; \sigma^2 = 1, \theta = 5, \nu )$. Following the model in \eqref{eq:1}-\eqref{eq:3}, the binary treatment $\bA$ and continuous outcome $\bY$ are generated as follows,
\begin{align}
 \bY &\sim Gaussian( \bA  + 0.55 \bZ_1 + 0.21 \bZ_2 + 1.17 \bZ_3 - 1.5 \bZ_4 + \bU, \sigma_e^2 = 0.1),  \label{eq:5} \\
    \bA &\sim Bernoulli(expit( -0.85 + 0.1 \bZ_1 + 0.2 \bZ_2 - 0.1 \bZ_3 - 0.7 \bZ_4 + c \bU)). \label{eq:6} 
    \end{align}
Each $\bZ_i$ is generated by independent standard normal random variables and is uncorrelated with $\bU$. The ATT in the above model equals $1$, the coefficient of $\bA$ in \eqref{eq:5}. 
In our simulations, we considered three different strengths of spatial confounding: strong, moderate, and weak, controlled by varying $c \in \{1.5,0.75,0.3\}$ in the propensity score model. We varied the spatial smoothness parameter, $\nu \in \{0.1, 0.5, 0.75, 1.5\}$ to assess the impact of varying spatial smoothness on the estimated causal effects. For each combination of $c$ and $\nu$, we generated 500 data sets. 
We evaluated the performance of our method under different scenarios. 
\subsection{Methods for comparison} \label{methods_comparison}
We compare our proposed method, referred to as ``RecoverU'', to three other methods. We consider the distance-adjusted propensity score matching (DAPSm) method proposed by \cite{dapsm}.
We further consider the generalized least squares estimate of $\beta$ in  the outcome model \eqref{eq:1} as the ``GLS'' estimate of ATT assuming that $U$ is uncorrelated to $A$. Additionally, we consider the standard IPTW estimator in the causal inference which does not incorporate any spatial information. We refer to this as the ``Na\"ive'' estimator. Finally, we compare all the methods above to the ``Gold standard'', which computes ATT using the doubly robust estimator based on the propensity score and the outcome model given $\bU$ were known. While the Gold standard considers $\bU$ to be known, our proposed method uses the partially recovered $\hat \bU_R$ as an approximation of the unknown $\bU$. 

For the Na\"ive and the DAPSm method, we fit the logistic propensity score model using observed covariates only. The ATT estimates for the Na\"ive method are obtained using the standard IPTW estimator described in \eqref{eq:att_iptw}. The 95\% normal confidence intervals are computed using the sandwich variance estimates. The DAPSm method is implemented using the ``DAPSm'' package in R. The method involves a tuning parameter $w$ which is chosen to be the optimal weight parameter using the package. 
The RecoverU method is implemented by following the detail steps given in Section \ref{Sec: Adjusting for unmeasured spatial confounders}. 
The ATT estimate is computed using equation \eqref{eq:att_dr_uhat} and its 
standard error was obtained using parametric bootstrapping. For each simulated data set, we obtained 500 bootstrap samples as follows; we assume the observed covariates to be known. The unmeasured spatial confounder, $\bU_{sim}$ is generated as a centered Gaussian random field  using the estimated spatial covariance matrix $\hat \Sigma_{\bU}$. The binary treatment $\bA_{sim}$ is generated using the fitted propensity score model given the covariates and $\bU_{sim}$. Finally, the outcome is generated using the fitted outcome model given the covariates,  $\bA_{sim}$ and $\bU_{sim}$. The bootstrap standard errors are computed as the standard deviation for each ATT estimator across the 500 bootstrap samples. The 95\% normal confidence intervals are constructed using the estimated ATT and the bootstrap standard error estimates. We evaluate the performance of the methods considered above under different scenarios using bias, variance, and the percent coverage of the ATT estimates.

\subsection{Results}
Through simulation studies, we found that the partially recovered spatial confounder $\hat \bU_R$ is highly correlated to the true $\bU$, which leads to a significant improvement while estimating the propensity score with observed covariates and $\hat \bU_R$. Hence, it achieves better covariate balance on the true $\bU$ than the Na\"ive method and the DAPSm method. For all the methods, we observe that the bias of ATT estimates decreases as the strength of spatial confounding decreases or as the spatial surface gets smoother.  RecoverU has the lowest bias and lowest variance as compared to the Na\"ive, GLS, and DAPSm methods across almost all the scenarios. The only exception is the GLS method has similar or slightly lower bias than RecoverU when the outcome model is correctly specified, the spatial surface is smooth and the spatial confounding is weak. Further, as the spatial surface gets smooth, the confidence interval coverage under RecoverU is around the nominal coverage except the case when both the outcome model and the propensity score models are misspecified. The following subsections provide a detailed description of these results.

\subsubsection{Improved propensity score estimates}

\begin{figure}[H]
  \centering\includegraphics[scale=0.71]{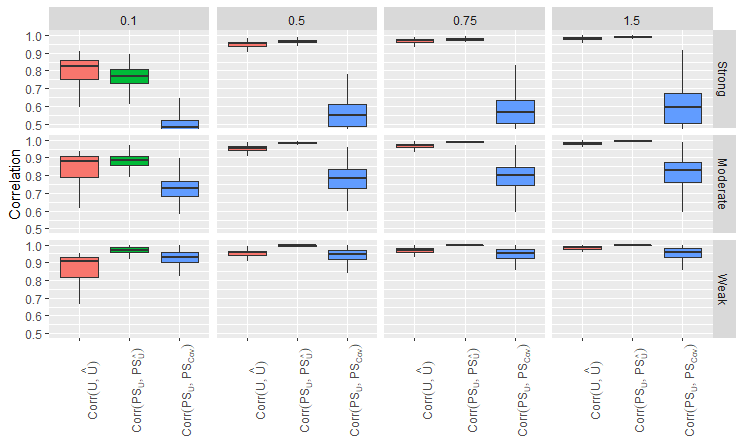}
  \caption{Correlations to assess quality of recovered unmeasured confounder. The boxplots summarize the different correlations over 500 simulations. The columns indicate the varying levels of spatial smoothness ($\nu = 0.1, 0.5, 0.75, 1.5$) and the rows indicate strength of spatial confounding (``Strong'', ``Moderate'', ``Weak'').}
  \label{fig:correct_correlations}
\end{figure}

We first consider the scenario in which both the propensity score model and the outcome model are correctly specified. The data is generated according to \eqref{eq:5} and \eqref{eq:6}. 
We begin by evaluating how the partially recovered $\bU$ compares to the true $\bU$ and how it helps improve the propensity score estimates. 
From Figure \ref{fig:correct_correlations}, 
we observe that $\bU$ and the partially recovered $\hat \bU_R$ are highly correlated, implying that the partially recovered $\bU$ provides a good approximation for the true unobserved $\bU$. The correlation between $\bU$ and $\hat \bU_R$ increases and is close to 0.95 as the spatial surface gets smoother $(\nu \geq 0.5)$. Denote by $PS_{\hat \bU_R}$, $PS_{\bU}$, and $PS_{Cov}$ the estimated propensity scores obtained by the RecoverU method, the Gold standard method,  and the Na\"ive method respectively.
Figure \ref{fig:correct_correlations} shows that the correlation between $PS_{\hat \bU_R}$ and $PS_{\bU}$ is significantly higher than the correlation between $PS_{Cov}$ and $PS_{\bU}$. Thus, the propensity score estimates obtained using our method are close to the true propensity scores and provide significant improvement over $PS_{Cov}$.

\begin{figure}[h]
  \centering\includegraphics[scale=0.71]{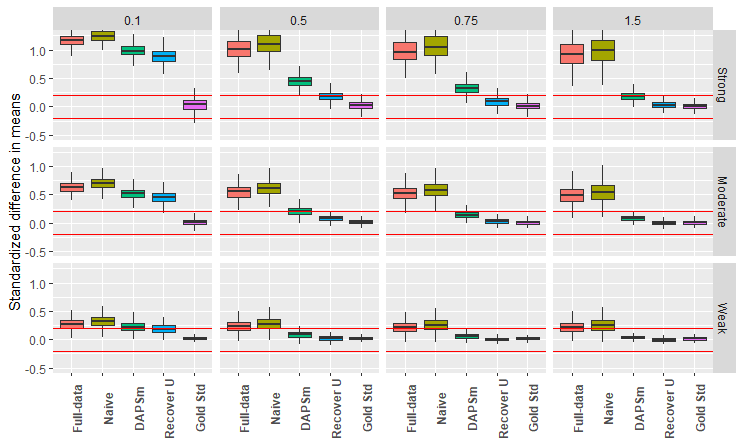}
  \caption{Standardized mean differences for $\bU$ across different methods. The boxplots summarize the distribution of SMD for $\bU$. The red solid lines represent the threshold of 0.2. }
  \label{fig:correct_balance_us}
\end{figure}

Next, we evaluate the different methods for the balance on the unobserved covariates. Figure \ref{fig:correct_balance_us} plots the distribution of standardized mean differences (SMD) of $\bU$ across different methods \citep{austin2009balance}. 
We consider a covariate to be balanced if the absolute SMD is within the 0.2 threshold \citep{lanza2013drawing}.
The ``Full-data'' refers to the balance on the covariates in the simulated data sets without any confounding adjustment. In the simulated data, $\bZ_4$ and $\bU$ are imbalanced across all scenarios. After propensity score adjustments, all methods achieve balance on the observed covariates and differ only with respect to balance on the unmeasured $\bU$.  
The Gold standard method uses the propensity scores conditional on the true $\bU$ to adjust for the imbalance. Thus, the absolute SMD between the treated and control group is always within the threshold, and balance is achieved on $\bU$. The Na\"ive method does not incorporate any spatial information and hence does not achieve balance on the unmeasured $\bU$. The RecoverU and DAPSm incorporate spatial information and improve balance on $\bU$ as the spatial surface gets smoother $(\nu \geq 0.5)$. Moreover, RecoverU results in smaller SMD on $\bU$ than DAPSm across all scenarios, thus achieving better balance on $\bU$ than DAPSm. 

\subsubsection{Effect of spatial confounding and spatial structure}
\label{effect_of_spatial_confounding_spatial_structure}

\begin{figure}[h]
  \centering\includegraphics[scale=0.53]{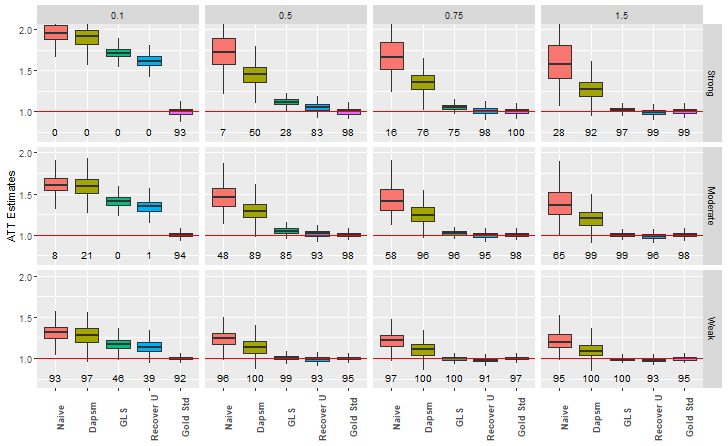}
  \caption{ATT estimates under a correctly specified outcome and propensity score model with non-spatial covariates. The boxplots summarize the distribution of ATT estimates. The solid line at $1$ represents the true ATT estimate. The percent coverage for 95\% confidence intervals is given below the boxplots.}
  \label{fig:correct_bias}
\end{figure}

\begin{figure}[H]
  \centering\includegraphics[scale=0.5]{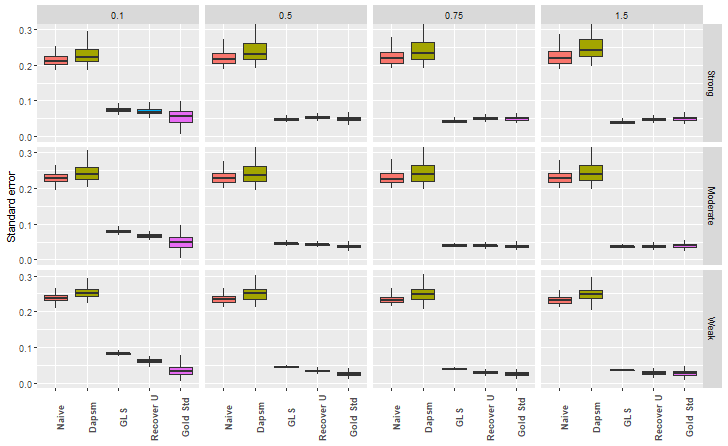}
  \caption{Standard error estimates for ATT under a correctly specified outcome and propensity score model with non-spatial covariates.}
  \label{fig:correct_SE}
\end{figure}

We assess the impact of the strength of spatial confounding and the impact of varying spatial smoothness on the causal estimates when the propensity score model and the outcome model are correctly specified. 
In Figure \ref{fig:correct_bias}, we observe that the bias in the ATT estimates highly depends on the strength of spatial confounding, with stronger confounding resulting in higher values of bias.  
When the spatial surface is rough ($\nu = 0.1$), all methods are biased; however, as the spatial surface gets smoother ($\nu \geq 0.5$), bias in estimates reduces. RecoverU provides smaller bias and smaller variance estimates compared to the Na\"ive, DAPSm, and GLS methods across almost all settings. The only exception is the GLS has similar or slightly lower bias when the spatial confounding is weak and the spatial surface is very smooth. As $\nu$ increases, the bias of RecoverU gets closer to the Gold standard, and the confidence interval coverage is around the nominal coverage of 95\%. The Na\"ive and the DAPSm method either have a coverage lower than RecoverU or have a coverage higher than the nominal coverage of 95\%. From Figure \ref{fig:correct_SE}, we can see that these two methods have significantly higher variance estimates compared to RecoverU across all settings, resulting in wider confidence intervals. The over-coverage problem is evident considering the higher than nominal coverage values even under large bias.

\subsubsection{Effect of model misspecification} \label{misspecification}

\begin{figure}[h]
  \centering\includegraphics[scale=0.55]{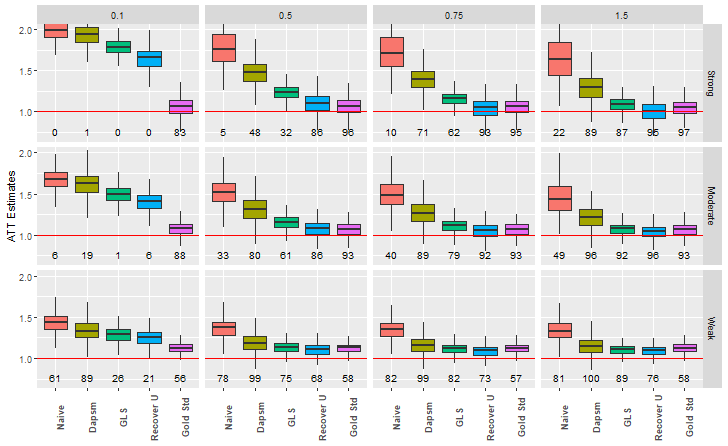}
  \caption{ATT estimates under a correctly specified propensity score model and misspecified outcome model with non-spatial covariates.}
  \label{fig:misspec_bias}
\end{figure}

\begin{figure}[h]
  \centering\includegraphics[scale=0.55]{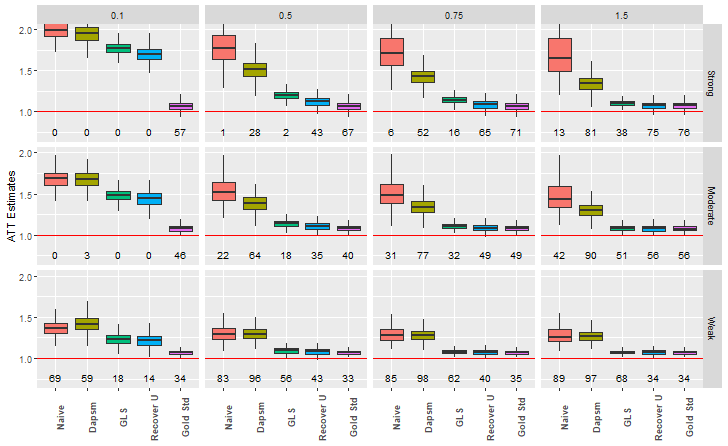}
  \caption{ATT estimates under misspecified outcome and misspecified propensity score model with non-spatial covariates.
}
  \label{fig:omitted_bias}
\end{figure}

To examine the robustness of the above methods to model misspecification, we consider the following two scenarios. First, only the outcome model is misspecified. Specifically, an additional interaction term $0.75 \bZ_3 \bZ_4$ is included in the true outcome model \eqref{eq:5}, which is omitted later on while fitting the outcome model. Second, the observed confounder $\bZ_4$ is omitted from both the outcome and the propensity score models while fitting these models; thus, both models are misspecified. The results are summarized in Figure \ref{fig:misspec_bias} and Figure \ref{fig:omitted_bias} respectively.

Under both scenarios, RecoverU has the lowest bias compared to all the other methods across all combinations of spatial smoothness and confounding, although the bias is relatively larger than the case when both the outcome and propensity score model are correctly specified in Section \ref{effect_of_spatial_confounding_spatial_structure}. 
In Figure \ref{fig:misspec_bias}, when only the outcome model is misspecified, as $\nu$ gets larger, the coverage of RecoverU is close to nominal except in the weak confounding case. DAPSm, Na\"ive and GLS still show relatively high bias and incorrect coverage. In Figure \ref{fig:omitted_bias} when both models are misspecified, as expected RecoverU is no more robust and has lower than nominal coverage despite low bias.  However, it still has better performance than the other three methods.


In both scenarios, the Gold standard estimate is also based on the misspecified models but considers $\bU$ is known. Comparing our method to the Gold standard helps us evaluate the capability of our method to recover the information about the missing confounders under model misspecification. Again, we observe that even under model misspecification, as the spatial surface gets smoother, the performance of RecoverU is similar to the Gold standard in terms of bias and coverage. We can thus conclude that the partially recovered $\bU$ is not affected even under outcome model misspecification. Also, since all the covariates are non-spatial, the missing terms in the misspecified model do not have any spatial structure to them and information on these terms cannot be recovered by the spatial methods. In the next set of simulations, we assess the impact of model misspecification when the missing terms have a spatial structure.    
\subsubsection{Impact of spatial vs. non-spatial confounders under model misspecification}


\begin{figure}[H]
  \centering\includegraphics[scale=0.55]{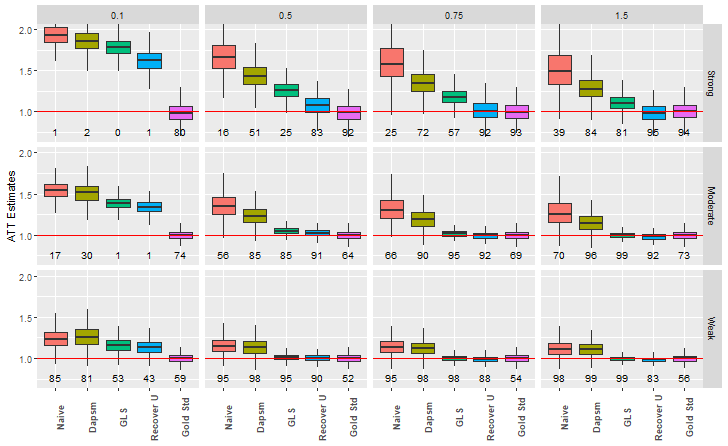}
  \caption{ATT estimates and coverage under a correctly specified propensity score model and misspecified outcome model with spatial covariates.
 }
  \label{fig:misspec_spatial_bias}
\end{figure}

\begin{figure}[H]
\centering\includegraphics[scale=0.55]{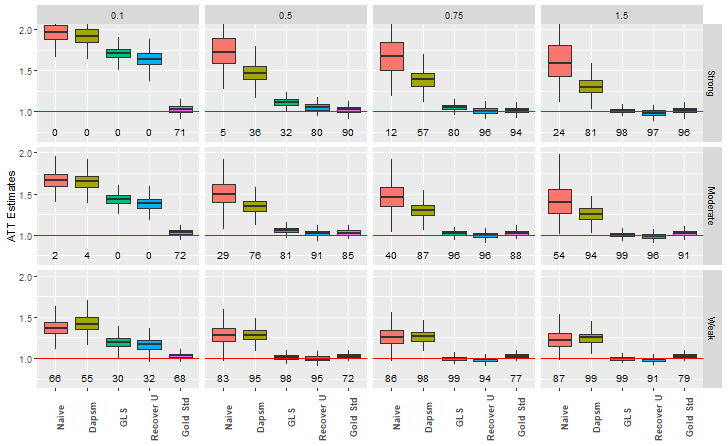}
  \caption{ATT estimates and coverage under a misspecified propensity score and misspecified outcome model with spatial covariates.
}
  \label{fig:omitted_spatial_bias}
\end{figure}

In all the above scenarios, the observed confounders $\bZ_i's$ are independent normal random variables. However, a more realistic situation in spatial causal inference would be when some of the covariates vary spatially. To this end, we re-consider the above scenarios under a set of spatial and non-spatial confounders. We consider two of the four confounders ($\bZ_1$ and $\bZ_2$) to be independent standard normal random variables. The rest ($\bZ_3$ and $\bZ_4$) are generated as centered Gaussian random fields with Mat\'ern covariance functions $C(h; \sigma^2 = 1, \theta = 3, \nu = 0.5)$, and $C(h; \sigma^2 = 1, \theta = 7, \nu = 0.5)$, respectively. Again, $\bZ_i$s are uncorrelated to $\bU$. Figure \ref{fig:misspec_spatial_bias} shows the results under a misspecified outcome model while Figure \ref{fig:omitted_spatial_bias} shows the results when both models are misspecified. 

Under misspecified models, when the missing terms have a spatial structure, RecoverU still has the lowest bias and lowest variance as compared to the Na\"ive and DAPSm across all settings. RecoverU outperforms GLS in most cases except when the spatial confounding is weak, and the spatial surface is very smooth. 
Since the missing interaction term and the omitted variable have a spatial structure, GLS and RecoverU can recover the information on these missing terms along with the unmeasured spatial confounder $\bU$. Thus, the bias of these methods here is much lower than with non-spatial covariates in Section \ref{misspecification}. Besides, the estimate of the Gold standard is biased, since it is based only on the true $\bU$ and does not incorporate information on the missing terms. Hence, RecoverU and GLS have higher coverage than the Gold standard in this case.

\section{Case study: Impact of SCR/SNCR technology on ambient ozone concentrations under unmeasured spatial confounding}
\label{Sec: Case study}
Ground-level ozone is a harmful pollutant; both short-term and long-term exposure to ozone is associated with increased risk of respiratory infections, cardiovascular harm, and increased mortality \citep{zhang2019ozone}. Ground-level ozone is not emitted directly into the air but is created when oxides of nitrogen $(NO_x)$ combine with volatile organic compounds (VOC) in the presence of heat and sunlight. Regulatory actions are necessary to reduce emissions of $NO_x$ and VOCs to improve air quality and reduce ground-level ozone. In the United States, electric utility power plants alone are responsible for a significant proportion of $NO_x$ emissions. Selective catalytic reduction (SCR) and Selective non-catalytic reduction (SNCR) emission control technologies are known to be most effective in reducing $NO_x$ emissions. However, does this reduction in $NO_x$ from installing SCR/SNCR technologies lead to a reduction in ambient ozone?



We use the power plant emissions data set publicly available on the Harvard dataverse to demonstrate the benefits of using our method.  The data was first analyzed in \cite{dapsm} using the DAPSm method. It consists of 473 power generating facilities powered by either coal or natural gas during June - August 2004.
A power plant can consist of multiple energy-generating units. The energy used at a facility, measured by heat input, can be used to determine the size of the facility. A power plant is considered to be treated ($A_i = 1$) if at least 50\% of its heat input is used by the energy generating units with at least one SCR or SNCR technology installed. A power plant is considered a control ($A_i = 0$) if it consists of facilities with some other type of $NO_x$ control technology installed, which is regarded to be less efficient than SCR/SNCR. 
For each power plant $i$, the response $Y_i$ is the $4$th highest daily ozone level averaged over all the ozone monitors within $100$ km of the power plant. The data includes $p=18$ observed confounders $\bZ_i$ describing the characteristics of the power plant and the area surrounding the power plants. $\bU$ represents spatial variables such as the pre-treatment $NO_x$ levels and the volatile organic compounds, which are known to be confounders but are unmeasured. We implement the Na\"ive, DAPSm, GLS, and RecoverU methods under unmeasured spatial confounding to estimate the effect of installing SCR/SNCR technologies versus alternatives on ambient ozone concentrations.

\subsection{Results} 

We first assess the covariate balance across the treated and control group using SMD. The threshold is set to 0.15, following the threshold used in \cite{dapsm}. Table \ref{tab:cov bal table} displays the variables included in the outcome model and the propensity score model. The propensity score model is fitted using logistic regression. All the methods are implemented as described in Section \ref{methods_comparison}. ``Full-data" refers to the SMD between the treated and control group without any confounding adjustment. We observe that 10 of the 18 observed covariates and $\hat \bU_R$ are unbalanced. After covariate adjustment using the Na\"ive method (non-spatial), two of the observed covariates and $\hat \bU_R$, still remain unbalanced. Balance is achieved on all variables using the DAPSm and RecoverU methods, which incorporate spatial information. The average absolute SMD using RecoverU is smaller than the average absolute SMD using DAPSm method.

\begin{table}[h]
\caption{\label{tab:cov bal table} The table displays the Standardized mean differences (SMD) for power plant characteristics and area level characteristics across different methods.}
\vspace{0.5cm}\
\begin{tabular}{llrrrr}
\hline \hline
                         &         & \multicolumn{4}{l}{Standardized mean differences (SMD)} \\
Variable                 & Type    & Full-data      & Naïve      & RecoverU     & DAPSm      \\
\hline
\% Operating capacity    & Contin. & 0.026          & -0.029     & -0.073       & 0.069      \\
ARP Phase 2              & Binary  & 0.253          & -0.026     & -0.033       & 0.149      \\
4th Max Temperature      & Contin. & 0.158          & 0.025      & 0.004        & 0.004      \\
\% Urban                 & Contin. & 0.131          & -0.036     & -0.033       & -0.062     \\
\% White                 & Contin. & -0.171         & 0.153      & 0.077        & -0.042     \\
\% Black                 & Contin. & -0.368         & -0.071     & -0.010       & -0.016     \\
\% Hispanic              & Contin. & 0.389          & 0.066      & -0.001       & 0.100      \\
\% High School           & Contin. & -0.244         & 0.057      & 0.044        & 0.087      \\
Median Household Income  & Contin. & 0.107          & -0.019     & 0.029        & -0.087     \\
\% Poor                  & Contin. & 0.105          & -0.034     & -0.075       & 0.101      \\
\% Occupied              & Contin. & 0.031          & -0.046     & -0.007       & 0.028      \\
\% 5-year residents      & Contin. & 0.083          & -0.008     & -0.047       & 0.012      \\
Median House Value       & Contin. & 0.339          & -0.121     & -0.013       & -0.083     \\
Heat Input               & Contin. & 0.156          & -0.061     & -0.097       & 0.143      \\
Population / square mile & Contin. & 0.105          & -0.165     & -0.069       & -0.032     \\
Gas facility             & Binary  & 0.482          & -0.002     & -0.027       & -0.035     \\
Small size facility      & Binary  & 0.196          & 0.000      & 0.036        & 0.016      \\
Medium size facility     & Binary  & -0.069         & -0.003     & -0.041       & -0.017     \\
$\hat \bU_R$                     & Contin. & 0.419          & 0.225      & -0.087       & -0.018     \\
\hline
\end{tabular}
\end{table}

\begin{figure}[h]
  \centering\includegraphics[scale=0.32]{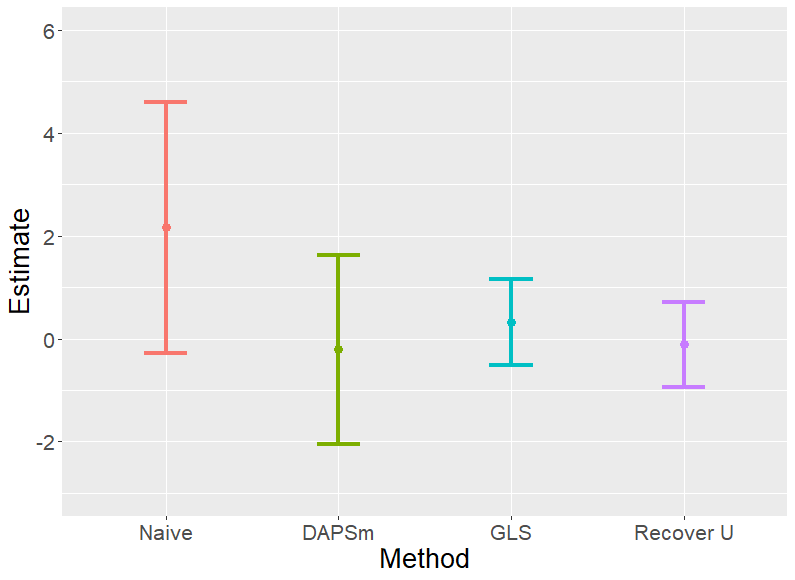}
  \caption{ATT estimates and 95\% confidence intervals for the effect of SCR/SNCR emission control technology on fourth maximum ozone concentration (in ppb).}
  \label{fig:ATT full m}
\end{figure}

Next, we evaluate the effect of installing SCR/SNCR technologies versus alternatives on ambient ozone concentrations. The point estimate of ATT for the RecoverU method is -0.11 parts per billion (ppb) and the corresponding 95\% CI is $(-0.94, 0.71)$. The interval however includes zero, suggesting that SCR/SNCR technologies do not reduce ambient ozone more than alternative technologies. 
We look at the ATT estimates and the 95\% confidence intervals across different methods. The results are shown in Table \ref{tab:ATT full m} and Figure \ref{fig:ATT full m}. 

\begin{table}[h] 
\centering
\caption{\label{tab:ATT full m} ATT estimates and 95\% confidence intervals for the effect of SCR/SNCR emission control technology on fourth maximum ozone concentration (in ppb).}
\vspace{0.5cm}\
\begin{tabular}{lrrr}
\hline \hline
         & \multicolumn{3}{r}{95\%  Normal Confidence Intervals} \\  
Method  & Estimate       & Lower bound             & Upper bound       \\ \hline
Naïve   & 2.16   & -0.28                      & 4.60              \\
DAPSm    & -0.21  & -2.05                     & 1.63              \\
GLS      & 0.32   & -0.51                     & 1.16              \\
RecoverU  & -0.11  & -0.94                     & 0.71              \\ \hline
\end{tabular}
\end{table}

The Naive method provides a point estimate of 2.16, which is positive and away from zero, and the 95\% CI for ATT estimate is (-0.28, 4.60). However, after incorporating spatial information, the interval shifts downwards, and the point estimates using GLS, DAPSm, and RecoverU are all very close to zero. All the methods, however, include zero in the confidence interval. The confidence intervals obtained using the Naive and the DAPSm method are wider compared to confidence intervals obtained using GLS and RecoverU methods. One of the benefits of using RecoverU lies in narrower confidence intervals compared to alternatives. 

From the simulation study results, we found that when $\nu > 0.5$, the ATT estimates provided by RecoverU have lower bias and lower variance than the Na\"ive and the DAPSm. For the power plant data, the smoothness parameter of the unmeasured confounder was estimated to be 0.7, thus the spatial surface is smooth and if either model is correctly specified and the causal assumptions are met, the ATT should be well estimated. Even though the results from GLS and RecoverU are similar here, RecoverU is still preferred since it is more robust to the strength of spatial confounding, the smoothness of unmeasured confounder, and the model misspecification. 


In the simulation study, the actual ATT estimates were known, and it was possible to show that our method achieves balance on the unmeasured confounder. However, in the case of the power plant data, the actual ATT estimates and the unmeasured confounders are unknown. Thus, to demonstrate that our method adjusts for the unmeasured spatial confounders, we carry out additional analysis with the power plant data. 
This analysis considers the spatially varying covariates with the largest unadjusted SMD, namely, \% Hispanic, \% Black and Median house value (MHvalue). Without any propensity score adjustments, these three variables are unbalanced in the treatment and control groups and can be considered as potential confounders. 
Omitting these confounding variables from the analysis can result in biased ATT estimates.

\begin{figure}[h]
  \centering\includegraphics[scale=0.4]{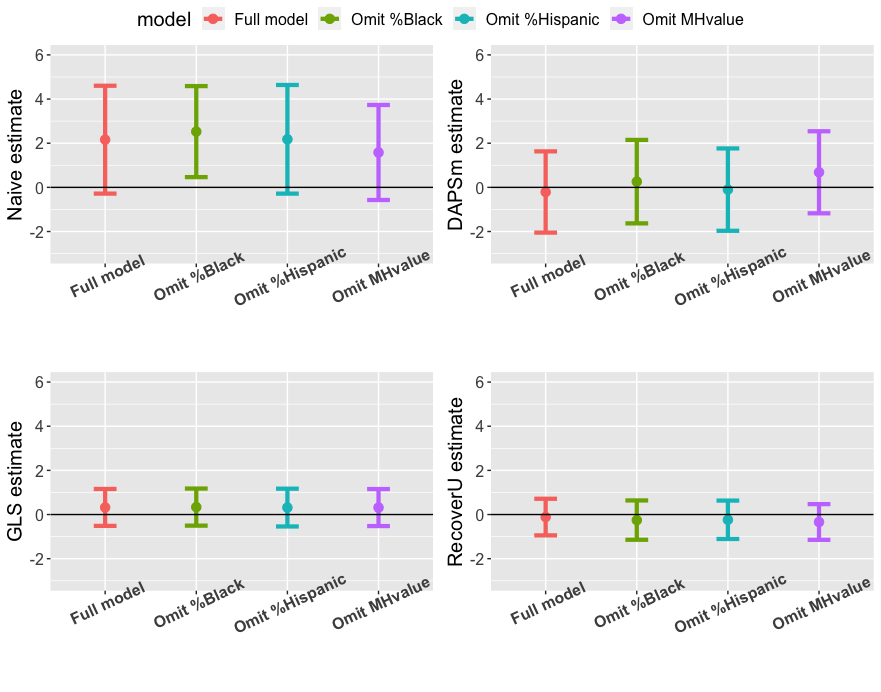}
  \vspace{-0.8cm}
  \caption{ATT estimates and 95\% confidence intervals for the effect of SCR/SNCR emission control technology on fourth maximum ozone concentration using the full model and by omitting observed confounding variables.}
  \label{fig:ATT omitted vars}
\end{figure}

We omit these variables one at a time and assess the impact of these missing spatial confounders on the ATT estimates. We assess if covariate balance is achieved on the omitted variable. We also compare the ATT estimates obtained from the omitted variable models to the ATT estimates listed in Table \ref{tab:ATT full m} using the complete set of covariates. 
The results for the ATT estimates and 95\% CI corresponding to the full model (i.e. model with full set of covariates) and the omitted variable models are shown in Table \ref{tab:ATT_omittedvars} and Figure \ref{fig:ATT omitted vars}.

When the \% Hispanic variable is omitted from the outcome and propensity score model, all methods still achieve balance on \% Hispanic and all other covariates.
When \% Black is omitted from the analysis, the Na\"ive method fails to balance on the \% Black and \% White variables. However, the DAPSm and RecoverU method achieve balance on \% Black and all the other covariates.
On the other hand, when the Median house value variable is excluded from the analysis, none of the methods achieve covariate balance on this variable. 

From Figure \ref{fig:ATT omitted vars}, we observe that when the \% Black variable is omitted from the analysis, the 95\% CI for the Naive method does not include zero anymore. The CI for the Naive method is positive, indicating that installing SCR/SNCR technology versus alternatives is associated with an increase in ambient ozone, which is contradictory to our knowledge. In this scenario, the Na\"ive method and the spatial methods give different results. All the spatial methods still include zero in the CI even when \% Black is omitted from the analysis.  
We also observe that the point estimates and CI for the DAPSm method shift upwards when the \% Black or the Median house value variables are omitted from the analysis. 
The ATT estimates using the RecoverU and GLS method are similar across all the omitted variable models. The shift in point estimates and 95\% CI is the least for the RecoverU and GLS methods as compared to other methods.

\begin{table}[h] 
 \caption{\label{tab:ATT_omittedvars} ATT estimates and 95\% confidence intervals for the effect of SCR/SNCR emission control technology on fourth maximum ozone concentration using full model and by omitting observed confounding variables.}
 \centering
\vspace{0.5cm}\
\begin{tabular}{llrrr} \hline \hline
         & \multicolumn{4}{r}{95\%  Normal Confidence Intervals} \\
                     & Method  & Estimate  & Lower bound  & Upper bound \\ \hline 
Full model            & Naïve     & 2.16       & -0.28          & 4.60        \\
                     & DAPSm   & -0.21 & -2.05           & 1.63        \\
                     & GLS       & 0.32 & -0.51           & 1.16        \\
                     & RecoverU & -0.11 & -0.94           & 0.71        \\ \hline
Omitting \% Hispanic & Naïve     & 2.18  & -0.28          & 4.64        \\
                     & DAPSm    & -0.10  & -1.97         & 1.76        \\
                     & GLS       & 0.32 & -0.54           & 1.17        \\
                     & RecoverU & -0.24 & -1.11          & 0.63        \\ \hline
Omitting \% Black    & Naïve    & 2.53 & 0.46             & 4.59        \\
                     & DAPSm    & 0.26 & -1.63            & 2.15        \\
                     & GLS      & 0.34  & -0.50           & 1.18        \\
                     & RecoverU  & -0.25 & -1.14          & 0.64        \\ \hline
Omitting Median      & Naïve    & 1.58 & -0.57            & 3.73        \\
house value          & DAPSm    & 0.68 & -1.18            & 2.54        \\
                     & GLS     & 0.32 & -0.52            & 1.15        \\
                     & RecoverU & -0.34  & -1.15         & 0.47        \\ \hline
\end{tabular}
\end{table}

From the above analysis, we can see that RecoverU provides similar results even when potential spatial confounders are omitted from the model. Omitting \% Hispanic or \% Black does not significantly impact the ATT estimates obtained using the RecoverU method. Our method is able to achieve covariate balance on these omitted variables. 
However, when the Median house value variable is omitted from the analysis, none of the methods can balance this variable. A possible explanation is that the Median house value variable is rough ($\nu = 0.38$) in space. Through simulation studies, we also observed that when spatial smoothness is rough ($\nu < 0.5$), the quality of the recovered spatial confounder may not be as good. Even though all the methods fail to achieve balance on the Median house value variable, the ATT estimates using the RecoverU method are similar to the ATT estimates under the full model. 


\section{Discussions}
\label{Sec: Discussions}


Unmeasured spatial confounding can be present in any models with spatially correlated residuals and poses a threat to valid causal inference. Existing methods in spatial statistics are based on outcome regression models whereas the ones in causal inference focus on using geographic proximity to adjust for unmeasured spatial confounders. Combining the tools in spatial and causal inference literature, we propose a novel method to incorporate spatial information in the propensity score model and obtain improved causal effect estimates with the DR estimator, which offers protection against either outcome or propensity score model misspecification. 

Through simulation studies, we demonstrate that the partially recovered $\bU$ is highly correlated to the true $\bU$. Incorporating the partially recovered $\bU$ in the propensity score model achieves balance on the true $\bU$ under most settings. This suggests that even though we only partially recover the unmeasured confounder, it can still help achieve balance on the missing confounder. Across almost all settings and even under model misspecification, the ATT estimates obtained using RecoverU have lower bias and lower variance than the Na\"ive, GLS, and DAPSm methods. Recovering the spatial structure of the unmeasured confounder provides better information than just using the spatial proximity of the units. Furthermore, under strong spatial confounding, the performance of RecoverU is significantly better than all the other three methods. When the spatial confounding is strong, $\bU$ is an important predictor of treatment assignment, and achieving balance on $\bU$ becomes essential for reducing bias.   


We implement the proposed method on the power plant emissions data to study the effectiveness of installing SCR/SNCR emission control technology on ambient ozone. The 95\% CI suggests that there is no significant difference between the SCR/SNCR technology and the alternative strategies. The ATT estimate of our RecoverU method has a much narrower confidence interval than the Na\"ive and DAPSm methods, which is consistent with our findings in the simulation study. Further, we find the RecoverU method provides similar results when potential spatial confounders are omitted from the model. This again supports that the proposed method is more robust under model misspecification.

Finally, in our method, we consider the random effect term to be a centered Gaussian random field with Mat\'ern covariance function. We use the Mat\'ern covariance structure since it is known to be flexible and can capture most spatially structured processes. Assumptions of isotropy and stationarity may not always be satisfied. If the spatial residual map shows signs of non-stationarity or anisotropic behavior, one can select a more suitable covariance structure instead of Mat\'ern covariance to model the spatial random effect term. Then, we can apply it to our proposed RecoverU method to obtain the causal effect estimates.


\bibliography{refs}

\end{document}